\documentclass{blois}

\def\Journal#1#2#3#4{{#1} {\bf #2}, #3 (#4)}

\def\RPP{{\em Rept. Prog. Phys.}}
\def\JPG{{\em J. Phys.} G}

\def\ra{\rightarrow}
\def\nutau{\nu_\tau}
\def\PoT{PoT.}

\newcommand{\Photo}{}

\begin{document}
\vspace*{4cm}
\title{The SHiP/NA67 experiment at the ECN3 high-intensity beam facility at the CERN SPS}

\author{M. Climescu on behalf of the SHiP Collaboration}

\address{Department of Physics and Astronomy, Ghent University,\\
B-9000 Ghent, Belgium}

\maketitle\abstracts{
  The Search for Hidden Particles (SHiP/NA67) is a general-purpose, high-intensity beam dump experiment approved in 2024 for the future exploitation of the ECN3 experimental hall at the CERN Super-Proton-Synchrotron in conjunction with the new Beam Dump Facility (BDF). It will collect $6\times10^{20}$ protons on target over $\sim$15 years of operation. SHiP is designed to probe the largely underexplored domain of feebly interacting particles with masses in the $\mathcal{O}(100~\mathrm{MeV})$ to few-GeV range, providing leading sensitivity to most models predicting particles within this range, notably heavy neutral leptons, dark photons, dark scalars, axion-like particles and light dark matter. The intense flux of neutrinos of all flavours produced in the dump additionally enables a rich Standard Model and neutrino-physics programme with notably $\mathcal{O}(10^3)$ $\nu_{\tau}$ per year of operation, thus bringing forward a study of $\nu_{\tau}$ phenomenology. This contribution summarises the physics motivation, the experimental concept and the detector subsystems, and outlines the expected sensitivity and timeline.}

\section{Physics motivation}

The Standard Model (SM), while very consistent, is unable to explain a number of well-established observations such as the smallness and origin of neutrino masses, the baryon asymmetry of the Universe or the nature of dark matter. These point unambiguously to physics beyond the Standard Model (BSM) which have escaped detection until now. They may be \emph{too heavy} to have been produced directly at existing machines, motivating searches at the high-energy frontier through facilities such as HL-LHC or FCC~\cite{deBlas:2025gyz}. Alternatively, it may be \emph{too weakly coupled}: light new states which interact with the SM would have evaded searches not because of their mass but because of their feeble interaction strength. Such feebly interacting particles (FIPs) provide powerful avenues to resolve the aforementioned issues with the SM~\cite{shipphys}. High intensity searches, where large production rates compensate for small couplings, are the appropriate strategy to explore this parameter space.

\section{Searching for FIPs at a beam dump}

A FIP which interacts with the SM inherits a decay width proportional to the square of its (small) coupling. For the couplings and masses of interest this places the laboratory decay length in the range $c\tau \sim 10^{-3}$--$10^{3}~\mathrm{m}$. The optimal experiment therefore combines: (i) a high-intensity beam to produce the rare states; (ii) a long, instrumented decay volume in which a FIP can decay back into visible SM particles; and (iii) high-precision scattering detectors and a decay spectrometer able to reconstruct the very rare decays through their final states while suppressing backgrounds to a negligible level. Proton beam dumps naturally provide these conditions. The interaction of an intense proton beam with a thick, high-$Z$ target maximises the production of heavy-flavour hadrons, which are copious sources of potential FIPs. The dominant challenge is the large muon flux emerging from the dump, which must be controlled together with the induced neutrino background in order to reach the zero-background conditions required for robust discovery conditions.

\section{The SHiP experiment at the Beam Dump Facility}

\begin{figure}[t]
\centerline{\includegraphics[width=\linewidth]{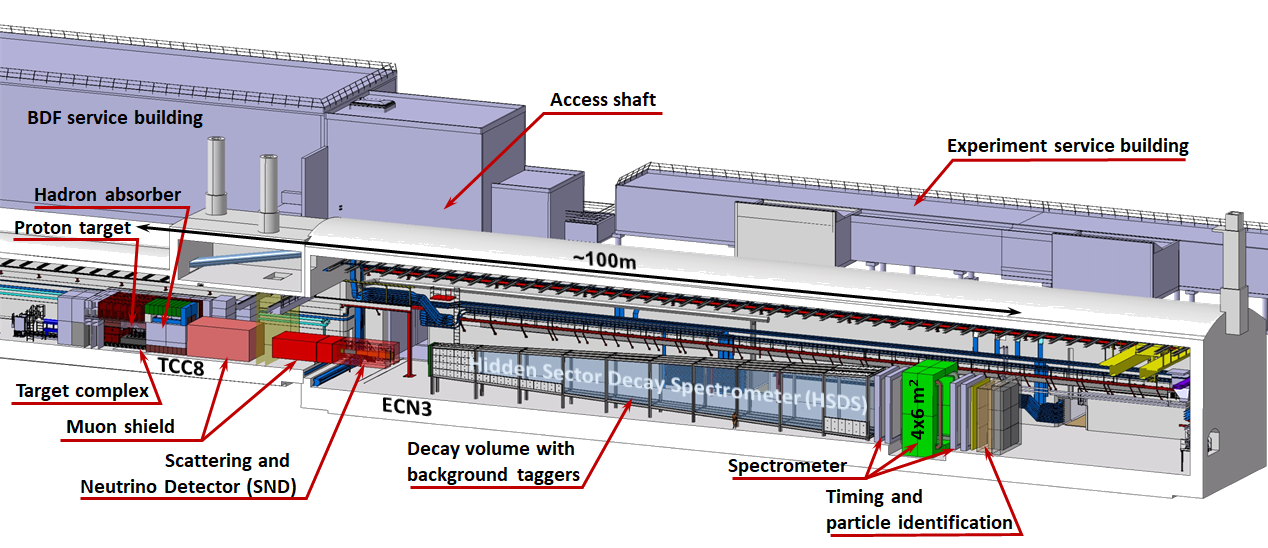}}
\caption[]{Overview of the BDF/SHiP experiment in the ECN3 hall. The proton beam comes in from the left and interacts with the target; the active muon shield deflects the residual muon flux, the background tagger systems are used to ensure a zero-background environment which allows for FIP observations in the downstream spectrometer, timing and calorimeter system.}
\label{fig:layout}
\end{figure}

The Beam Dump Facility (BDF) and the SHiP/NA67 experiment will perform these high-intensity searches after Long Shutdown~3, as part of the High-Intensity ECN3 (HI-ECN3) project~\cite{ecn3,ship2025}. SHiP will utilise $\mathcal{O}(4\times10^{19})$ $400~\mathrm{GeV}/c$ protons on target (\PoT) per year, accumulating $6\times10^{20}~\PoT$ over the lifetime of the programme. This unprecedented intensity yields fluxes $\mathcal{O}(10^{18})$ charmed hadrons and $\mathcal{O}(10^{16})$ beauty hadrons, giving SHiP a globally unique reach: for a generic FIP model in the $\mathcal{O}(100~\mathrm{MeV})$--few-GeV mass range, SHiP is expected to provide world-leading or record sensitivity. The experiment is sensitive to all benchmark portals -- HNLs, dark photons, dark scalars, ALPs and both elastic and inelastic light dark matter, as well as millicharged particles~\cite{ship2025}. The layout of the facility and the experiment is shown in Fig.~\ref{fig:layout}.

\section{Detector overview}

\subsection{Target and active muon shield}

The proton beam is dumped onto a tungsten target, so as to maximise the heavy-flavour production cross section; the very high beam intensity requires active cooling and a magnetised hadron absorber at the rear of the target complex. The extremely high ($\mathcal{O}(10^{11})$) muon flux emerging from the dump is the principal source of background and is suppressed by six orders of magnitude using a magnetic \emph{active muon shield} \cite{ship_ar2025}.

\subsection{Scattering and Neutrino Detector (SND@SHiP)}

The SND exploits the intense neutrino flux produced in the dump. It combines a dense high-granularity silicon-tungsten neutrino target with an iron-scintillator magnetised tracking calorimeter to identify (anti-)neutrino interactions of all flavours and reject backgrounds through topological and calorimetric selection. SHiP expects $\mathcal{O}(10^{4})$ $\nutau/\bar{\nu}_\tau$ interactions. This sample, together with high statistics in the other flavours, will enable the first determination of the structure functions $F_4$ and $F_5$ -- accessible only with tau neutrinos and never measured until now--, studies of parton distribution functions at low $x$ and a measurement of the strange-quark nucleon content via charm production in neutrino interactions. The detector also enables detection of light dark matter scattering off electrons and builds on the experience of SND@LHC~\cite{sndlhc}.

\subsection{Background taggers}

SHiP attains its zero-background goal through the use of a set of background tagger. The Upstream Background Tagger (UBT), placed at the entrance of the decay volume, tags residual muon background using a combination of straw trackers for position and scintillator tiles for timing. The Surrounding Background Tagger (SBT) encloses the helium-filled decay volume and tags charged particles as well as $\nu/\mu$ interactions in the vessel walls and in the helium. It is segmented into about 780 cells filled with $\sim$145{,}000~L of liquid scintillator, read out by wavelength-shifting optical modules (WOMs) coupled to SiPMs, achieving above $99\%$ efficiency and sub-1~ns timing resolution. Together with simple kinematic and impact-parameter selections, these systems reduce the dominant neutrino, muon deep-inelastic and muon combinatorial backgrounds to a negligible level while preserving signal efficiency.

\subsection{Hidden Sector Decay Spectrometer}

The decay spectrometer reconstructs the visible decay products of FIPs that decay inside the evacuated/helium-filled decay volume. It comprises:

\begin{itemize}
\item a \emph{spectrometer straw tracker} of four stations coupled to a large LHCb-style dipole magnet (nominal on-axis field $\sim$0.15~T,
  $0.6$--$0.8~\mathrm{T\,m}$). The very large detector ($4\times6~\mathrm{m}^2$) is expected to yield $120~\mu m$ spatial resolution. 
\item a \emph{timing detector} of three overlapping columns made of long scintillator bars read out by SiPMs, with a timing resolution $\leq$50~ps, used to suppress combinatorial background;
\item a \emph{calorimeter system} (electromagnetic and hadronic) based on the SplitCal concept~\cite{ecn3}. In addition to energy reconstruction with scintillator bars and SiPMs, dedicated high-precision layers measure the direction of electromagnetic showers, allowing the full reconstruction of events including neutral final states (such as $X\ra\gamma\gamma$) and providing particle identification for background rejection.
\end{itemize}

\section{Sensitivity and outlook}

\begin{figure}[t]
  \centering
\includegraphics[width=0.49\linewidth]{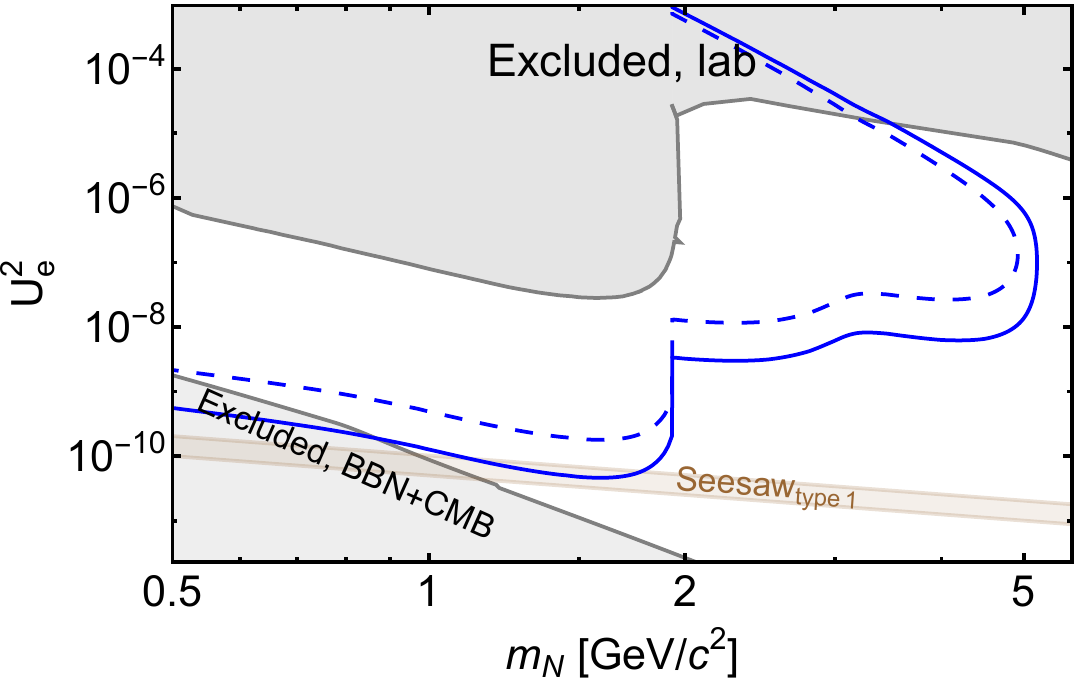}
\includegraphics[width=0.49\linewidth]{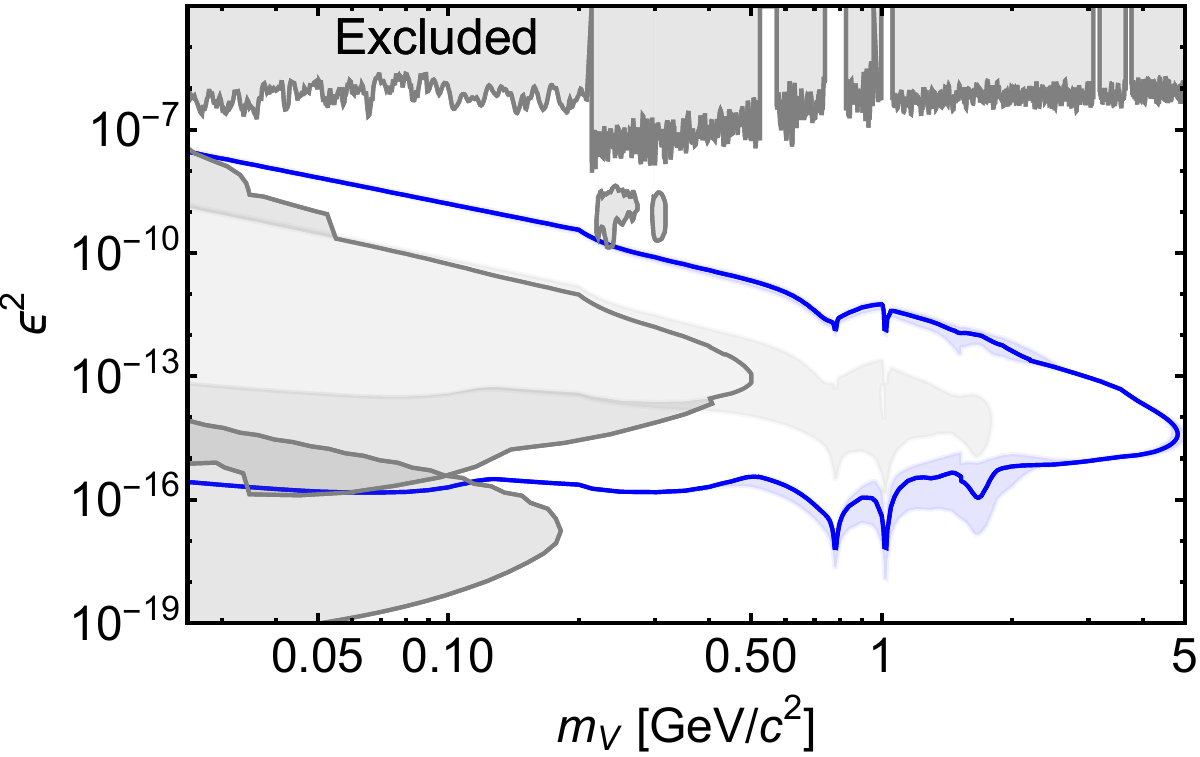}
  \caption[]{SHiP sensitivity to heavy-neutral-leptons with electron couplings [left] and dark-photons [right].}
\label{fig:sens}
\end{figure}

\begin{figure}[t]
\centerline{\includegraphics[width=\linewidth]{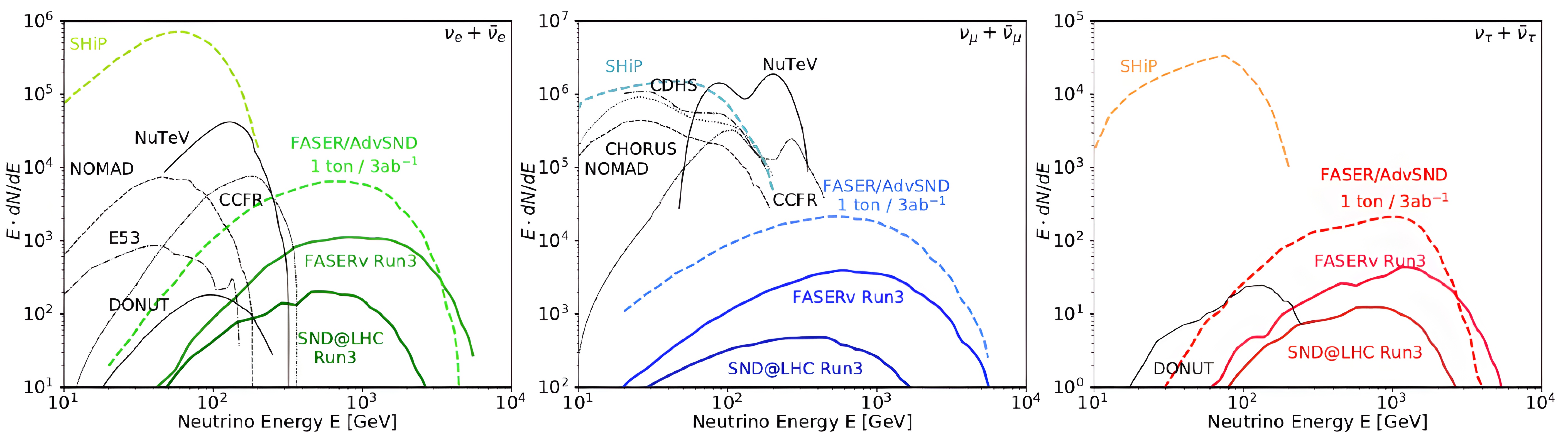}}
  \caption[]{SHiP neutrino yield and other existing and proposed experiments for comparison, adapted from \cite{beacham}.}
\label{fig:nu_sens}
\end{figure}

The combination of high intensity, a long decay volume, background tagging and complementary scattering and decay detectors gives SHiP discovery-level sensitivity across a broad set of FIP models, as illustrated in Fig.~\ref{fig:sens}. Over a large region of the $\mathcal{O}(100~\mathrm{MeV})$--few-GeV mass range SHiP probes couplings well beyond the reach of existing and other proposed experiments, operating in an essentially background-free regime. In parallel, the neutrino programme will exceed the integrated flux of all previous experiments in the $\mathcal{O}(10)~\mathrm{GeV}$ region for $\nu_e$ and, in particular, for $\nutau$ as shown in Fig.~\ref{fig:nu_sens}.

The collaboration is now in Technical Design Report phase with outputs expected within roughly two years. The BDF is being constructed and is foreseen to be commissioned with beam by 2033, with data taking to follow shortly thereafter. The data set collected before the subsequent long shutdown alone will already allow SHiP to set world-leading limits over much of its parameter space, with detector optimisation currently ongoing in preparation for the TDRs.

\section{Summary}

The energy frontier has so far revealed no new physics beyond the SM, strongly motivating a complementary exploration of the intensity frontier. SHiP/NA67 will have world-leading sensitivity to both explore this frontier directly as well as study all-flavour neutrino interactions. Approved in 2024 and to be built within the coming years at the CERN SPS, it will begin data taking in 2033.

\section*{Acknowledgments}

The author was supported by the Research Foundation -- Flanders (FWO) under grant \#12A4O26N, and thanks the organisers of the 37\textsuperscript{th} Rencontres de Blois for a stimulating conference.

\end{document}